%% file: paper.tex
\documentclass[11pt,a4paper]{llncs}

\usepackage{style}

\begin{document}
\input{acronyms.tex}


\title{\LARGE{Survey: Understand the challenges of Machine Learning Experts using Named Entity Recognition Tools}}


%
%
\author{\large{Florian Freund\orcidlink{0000-0002-7344-6869} \and Philippe Tamla\orcidlink{0000-0002-0786-4253} \and Matthias Hemmje\orcidlink{0000-0001-8293-2802}}}
\institute{\large{University of Hagen, Faculty of Mathematics and Computer Science\\58097 Hagen, Germany}\\
\email{\{florian.freund, philippe.tamla, matthias.hemmje\}@fernuni-hagen.de}}

%


%
%


\maketitle

\thispagestyle{firstpage}

\begin{abstract}
This paper presents a survey based on Kasunic's survey research methodology to identify the criteria used by Machine Learning (ML) experts to evaluate Named Entity Recognition (NER) tools and frameworks. Comparison and selection of NER tools and frameworks is a critical step in leveraging NER for Information Retrieval to support the development of Clinical Practice Guidelines. In addition, this study examines the main challenges faced by ML experts when choosing suitable NER tools and frameworks.
Using Nunamaker's methodology, the article begins with an introduction to the topic, contextualizes the research, reviews the state-of-the-art in science and technology, and identifies challenges for an expert survey on NER tools and frameworks. This is followed by a description of the survey's design and implementation. The paper concludes with an evaluation of the survey results and the insights gained, ending with a summary and conclusions.
\end{abstract}

\section{Introduction and Motivation}
\label{sec:introduction}

\ac{NER} tools, including libraries and frameworks, were introduced in the early 1990s \cite{cunningham2013getting} and have continued to evolve since then. Gudivada defined libraries and frameworks as follows:
\textit{``A software library is a set of functions that application can call, whereas a framework provides higher-level support in the form of some abstract design to speed up applications development''} \cite{gudivada_open-source_2018}.
Both libraries and frameworks are commonly used in \ac{NER} applications, depending on the specific requirements and complexity of the task at hand.
\ac{NER}, a sub-discipline of \ac{NLP}, plays a critical role in the extraction of knowledge from unstructured text, a particularly valuable task in healthcare where \ac{IO} is a persistent challenge \cite{savova_mayo_2010,gavrilov_feature_2020,liu_med-bert_2021,nawroth_emerging_2020}. \ac{IO} \cite{bawden_dark_2009} complicates the development of \acp{CPG} \cite{steinberg2011clinical,valika_second_2020}, as the search for evidence relies on vast amounts of unstructured text, such as clinical reports and findings of medical research \cite{valika_second_2020}. By converting unstructured text into structured data, \ac{NER} helps manage \ac{IO} in the medical domain \cite{nawroth_emerging_2020}. 
These data can be used to support \ac{IR} for the search for evidence as the basis for \acp{CPG} \cite{freund_towards_2023}. The use of \acp{CPG} can help reduce the risk for the patient in clinical decision-making \cite{steinberg2011clinical}.
Modern text analysis techniques have led to the development of \ac{ML}-based methods for \ac{NER} \cite{chieu_named_2003,florian_named_2003, mccallum_early_2003,peters_deep_2018}. These methods are highly efficient at processing unstructured natural language text \cite{konkol_named_2015}. The most commonly used \ac{ML} methods for \ac{NER} include supervised, unsupervised, and semi-supervised learning \cite{konkol_named_2015}. Supervised learning, which relies on manually annotated data to train models, is currently the most widely used method \cite{li_survey_2022,ariruna_dasgupta_classification_2016}. Unsupervised learning, on the other hand, uses statistical algorithms to identify patterns in unlabeled data \cite{ariruna_dasgupta_classification_2016}. Semi-supervised learning combines both approaches and requires only a small amount of annotated data \cite{ariruna_dasgupta_classification_2016}. The growing interest in \ac{ML}-based \ac{NER} has led to a significant increase in research activity and the number of available \ac{ML}-based \ac{NER} tools \cite{casey_systematic_2021}. Consequently, it can be challenging for users to keep up with the latest developments and stay informed about the state-of-the-art in this field.

Having described that \ac{NLP}, \ac{NER}, and \ac{ML} are essential techniques for effective \ac{IR} in the medical field, this research is now motivated by related research projects.
The \ac{RecomRatio} project, launched by Bielefeld University in 2018 to develop a computational method to rationalize recommendations, uses the medical literature to extract arguments for or against a particular medical treatment \cite{bielefeld_university_ratio_2017, nawroth_supporting_2020}. Arguments are made available in a knowledge base to support medical decisions.
\acp{CPG} can help reduce patient risk in medical decision making. However, domain experts face the challenge of \ac{IO} when developing \acp{CPG} \cite{valika_second_2020} since they need to use large amounts of unstructured text, such as clinical reports or medical research findings, as sources of evidence \cite{steinberg2011clinical}. 
To make the knowledge within these documents accessible, an automated analysis and visualization of specific \ac{NLP} features in natural language texts are essential, such as 
\ac{NER}, Entity Linking, or Relation Extraction \cite{li_neural_2021}.
The \ac{AI4H3} project builds upon the results of \ac{RecomRatio} and aims to support the transparency and explainability of medical decisions using \ac{AI} \cite{ftk_artificial_2020}. For this purpose, \ac{AI4H3} proposes a layered architecture with a central hub called \textit{''KlinGard Smart Medical Knowledge Harvesting Hub``}. This hub serves as a central point for registering \ac{AI} modules that can be used for natural language text analysis. In addition to integrating various technologies, this hub architecture allows heterogeneous data and \ac{AI} modules to be integrated in a decentralized manner.
The \ac{CIE} project deals with the hub architecture of \ac{AI4H3} \cite{tamla_cie_2023}. This concerns, among other things, the provision of cloud-based resources (such as computing power and storage) for the automatic extraction of natural-language texts using \ac{ML} techniques \cite{tamla_cie_2023}. These resources are intended to enable end-to-end \ac{NER} pipeline support in a cloud environment. 
To successfully use \ac{ML}-based \ac{NER} in knowledge domains, such as medicine, domain-specific knowledge is necessary for developing and training \ac{ML} models. \ac{ML}-based \ac{NER} could be used more widely for information extraction if domain experts could train and use \ac{NER} systems independently. \ac{FIT4NER}, also a project in the \ac{AI4H3} environment, aims to enable medical experts to use various \ac{AI}-based text analysis techniques \cite{freund_fit4ner_2023}.
For domain experts, the dynamic nature of \ac{NER} research presents several challenges. 
Firstly, \ac{NER} users need to compare various tools helping them to identify \ac{NLP} features, such as like \acp{NE} and Entity Relations, before deciding which solution is best suited for their tasks.
However, comparing different solutions can be challenging because \textit{``it remains difficult for \ac{NLP} practitioners to clearly and objectively %
identify what software perform(s) the best''} \cite{schmitt_replicable_2019}, 
along with determining which tools efficiently extract, analyze, and visualize \ac{NLP} features for effective \ac{IR} to support \ac{CPG} development.
Existing studies have used different datasets and presented their results heterogeneously, making it difficult to compare between them \cite{casey_systematic_2021}.
Secondly, \ac{NER} users face the challenge of selecting an appropriate tool for their specific task based on the comparisons aforementioned. This step \textit{``is critical in developing an \ac{NLP}-based application as it affects the accuracy of analysis tasks''} \cite{weiying_benchmarking_2019}.
Third, domain users often lack the computational and storage resources necessary to train high-quality \ac{NER} models in their knowledge domain. Although cloud computing could potentially address this issue, domain experts often lack the knowledge and experience necessary to effectively leverage this technology \cite{vouk_cloud_2008,chard_cost-aware_2015}.
The primary objective of this survey is to explore the challenges faced by \ac{ML} experts when \textbf{comparing} and \textbf{selecting} \ac{NER} tools for their projects. The findings from this research will be used to help non-experts make informed decisions when comparing and selecting \ac{NER} tools. To achieve this goal, the following \acp{RQ} have been defined and will be addressed in this work:
\textit{(\ac{RQ}1) How do \ac{ML} experts evaluate \ac{NER} tools, and which criteria are most important to them? (\ac{RQ}2) What primary challenges do \ac{ML} experts face when selecting a suitable \ac{NER} tool?}

This study is structured using the Nunamaker research method \cite{nunamaker_jr_systems_1990}, which consists of four phases: observation, theory building, system development, and experimentation. 
Chapter \ref{sec:observation} belongs to the observation phase and focuses on analyzing the current state of the art and related work to this research. Chapter \ref{sec:theorybuilding} is dedicated to survey modeling as part of the theory building phase and involves developing the survey questionnaire, which belongs to the system development phase. In Chapter \ref{sec:implementation}, the results of experiments with \ac{ML} experts are described as part of the experimentation phase. Finally, Chapter \ref{sec:conclusion} summarizes the study findings.

\section{State of the Art in Science and Technology}
\label{sec:observation}

This chapter focuses on the observation phase and introduces the background of this work and related research activities.
The objective is to identify and discuss \acp{RC} in the areas addressed in this article.
First, \ac{NER} and the challenges of dealing with various \ac{NER} tools are described. Second, a short introduction to cloud technologies is given. Finally, related studies are presented and discussed.

\ac{NER} is an \ac{NLP} technique that aims to extract \acp{NE} from unstructured text documents \cite{jehangir_survey_2023}. A \ac{NE} is a word or phrase that refers to a specific entity such as a person, place, or organization.
\ac{NER} is a crucial technique used in various applications, including \ac{IR} \cite{petkova_proximity-based_2007}, question answering systems \cite{barskar_approach_2012}, machine translation \cite{marton_transliteration_2014}, and social media analysis \cite{kim_weakly_2022}. In the medical domain, \ac{NER} plays a pivotal role in \acp{CDSS} and enables clinical information mining from \acp{EHR} \cite{pagad2022clinical}.
In recent years, \ac{NER} has seen significant progress due to the development of new techniques and models, including deep learning \cite{li_survey_2022}. These advancements have led to substantial improvements in the performance of \ac{NER} systems, making \ac{NER} one of the most extensively researched \ac{NLP} tasks today \cite{li_survey_2022}.
In \ac{NER}, there are different techniques available, including traditional, \ac{ML}-based, and hybrid approaches \cite{konkol_named_2015}. Traditional \ac{NER} approaches rely on methods that use manually created rules or are dictionary-based. Although these systems are often efficient and accurate, they are also limited by fixed rules or dictionaries and do not generalize well across different domains and languages \cite{konkol_named_2015}.
\ac{ML}-based approaches to \ac{NER} have gained popularity in recent years, mainly due to the availability of large annotated datasets and advancements in deep learning techniques \cite{li_survey_2022}. These approaches are capable of efficiently processing unstructured and large datasets and achieve superior results. Instead of relying on fixed rules or dictionaries, \ac{ML}-based \ac{NER} uses statistical models that learn to detect \acp{NE} from annotated data through a process of training and testing.
\ac{ML} techniques are divided into supervised, unsupervised, and semi-supervised learning \cite{ariruna_dasgupta_classification_2016}. Supervised learning \cite{ariruna_dasgupta_classification_2016} relies on manually annotated data to train a model, where the model learns to predict the labels of unseen data. Unsupervised learning \cite{ariruna_dasgupta_classification_2016}, on the other hand, relies only on statistical algorithms to detect patterns from unlabeled data. Semi-supervised learning \cite{ariruna_dasgupta_classification_2016} combines these two approaches by training a model with a small set of annotated data and using it to label a larger set of unlabeled data, thus improving the accuracy of the model.
In recent years, pre-training large language models such as BERT \cite{devlin_bert_2018}, GPT-2 \cite{radford_language_2019}, and RoBERTA \cite{liu_roberta_2019} on large corpora have shown remarkable improvements in \ac{NER} performance. These models are capable of achieving state-of-the-art performance on \ac{NER} tasks and can efficiently fine-tune on smaller datasets for domain-specific tasks. Although further improvements can be made, \ac{AI} advancements have already made significant progress in addressing complex \ac{NER} challenges.
The research field of \ac{NER} continues to evolve rapidly, with new and innovative tools being developed to address different challenges and use cases. Therefore, it is crucial for \ac{ML} experts to compare and evaluate the performance of available \ac{NER} tools and to select the one that best fits their specific task, such as training and fine-tuning \ac{ML} models on custom datasets. This study aims to gain insight into how such comparisons are conducted in practice and identify the challenges and factors that influence the decision-making process of \ac{ML} experts (\ac{RC}1).

\ac{AWS} launched in the early 2000s, pioneering the concept of cloud computing by offering scalable computing resources on demand as a service \cite{achar_cloud-based_2019}. This groundbreaking technology has since evolved into a widely available solution that offers vast amounts of computing resources at any given time. 
The availability of scalable and cost-effective cloud computing has revolutionized the field of \ac{AI} by providing a scalable and cost-effective platform for creating, training, and deploying \ac{AI} models. \ac{ML}-based \ac{NER} is one of the many \ac{AI} applications that have benefited from the cloud's capabilities \cite{tamla_cie_2023}.
The unprecedented growth of data has made it challenging to manage and analyze large amounts of information using local compute resources \cite{li_survey_2022}. To tackle this issue, leading providers such as \ac{AWS}, Microsoft Azure, and Google Cloud Platform offer cloud-based platforms at various levels of abstraction, including \ac{IaaS}, \ac{PaaS}, and \ac{SaaS} \cite{achar_cloud-based_2019}. 
These platforms provide the necessary computing resources and tools to store, process, and analyze massive amounts of data efficiently and cost-effectively.
Cloud-based \ac{ML} platforms not only provide computing power (\ac{IaaS}), but also offer a comprehensive suite of tools and services for data processing, model training, and deployment (\ac{PaaS}). These platforms make it easy to scale performance up or down as needed, even for demanding applications with real-time requirements.
Cloud providers offer not only cloud-based \ac{ML} platforms but also \ac{NLP} and \ac{NER} services. These services include pre-built models and \acp{API} that enable users to easily incorporate \ac{AI} functionality into their applications without requiring extensive expertise in the \ac{AI} domain  \cite{tamla_cie_2023}. 
To leverage cloud-based resources effectively, \ac{ML} experts must carefully evaluate which level of abstraction and which cloud-based services from which provider to use. Furthermore, utilizing cloud-based resources requires familiarity with the relevant technologies, including understanding their strengths, limitations, and best practices \cite{kim_cloud_2009}.
Although cloud technology offers many benefits, including scalability and cost effectiveness, there are legitimate concerns about privacy, security, and ethical implications, particularly in the medical field \cite{blohm_towards_2019}. As a result, \ac{ML} experts must carefully consider these factors and evaluate whether cloud-based resources can be used while still meeting regulatory and ethical standards.
In summary, cloud technology has rapidly evolved into a powerful platform for creating, training, and deploying \ac{AI} models. However, \ac{ML} experts face the challenge of determining whether and which cloud-based resources to deploy, requiring careful evaluation of factors such as scalability, cost, security, privacy, and ethical implications.
This study aims to conduct a survey of \ac{ML} professionals to uncover the key factors they consider when deciding whether to use cloud-based resources (\ac{RC}2).

In recent years, several scientific papers have compared and evaluated \ac{NER} tools for various application domains, such as formal and social media texts \cite{pinto_comparing_2016}, software documentation \cite{al_omran_choosing_2017}, historical texts \cite{won_ensemble_2018}, news sources \cite{weiying_benchmarking_2019,schmitt_replicable_2019}, and specific languages \cite{aldumaykhi_comparing_2022}.
Pinto et al. \cite{pinto_comparing_2016} conducted a study to compare and analyze the performance of multiple \ac{NLP} tools, including their effectiveness on formal and social media texts in four commonly used \ac{NLP} tasks, which include \ac{NER}.
Their findings suggest that it is a challenge \textit{``to select which one to use, out of the range of available tools''}, and \textit{``this choice may depend on several aspects, including the kind and source of text''} \cite{pinto_comparing_2016}.
Al Omran et al. \cite{al_omran_choosing_2017} conducted a comprehensive systematic review and experiments in 2017 to analyze the appropriate selection of an \ac{NLP} library for the analysis of software documentation. Their study focused on tokenization and part-of-speech tagging, which are essential tasks in the process of performing \ac{NER}.
Their findings underscored the criticality of selecting the right library, yet revealed that a small proportion of papers in the literature provide justification for their \ac{NLP} library choices. Based on their results, the authors strongly recommend that researchers carefully consider their options when comparing and selecting \ac{NLP} libraries and make informed decisions.
During a comparison of \ac{NER} tools for use with historical texts in 2018, Won et al. \cite{won_ensemble_2018} discovered that the \textit{``individual performance of each \ac{NER} system was corpus dependent''}. By combining various tools, they were able to achieve superior results without the need to translate historical texts into modern English.
In 2019, Weiying et al. \cite{weiying_benchmarking_2019} conducted a benchmarking study of \ac{NLP} tools for enterprise applications, which included standard \ac{NLP} tasks such as \ac{NER}. Their research highlights the importance of carefully selecting an appropriate \ac{NLP} library as a crucial step in the development of \ac{NLP} applications.
Schmitt et al. \cite{schmitt_replicable_2019} identified the challenges associated with selecting a \ac{NER} tools. They found that objectively comparing \ac{NER} tools is challenging due to the lack of replicable existing comparisons, and that research surveying \ac{NER} tools users about the difficulties of selecting and comparing these tools is rare.
Aldumaykhi et al. \cite{aldumaykhi_comparing_2022} recently conducted a comparative study of three \ac{NER} tools for analyzing Arabic texts. Through experimentation, they also found that combining these tools resulted in improved performance.
Jehangir et al. \cite{jehangir_survey_2023} examined the most relevant datasets, tools, and deep learning approaches currently used for \ac{NER} problem solving. Among other things, they discussed five different available tools utilized for \ac{NER}, such as spaCy, NLTK, and OpenNLP \cite{jehangir_survey_2023}. 
They found that each model or approach has its methodological advantages and disadvantages. For instance, Deep Learning offers benefits in terms of feature engineering and implementation complexity, while rule-based methods require significant manual effort for rule generation and are complex to implement \cite{jehangir_survey_2023}.
Additionally, they noted that combining various models or approaches could potentially yield superior results \cite{jehangir_survey_2023}.
This underscores the necessity of comparing \ac{NER} approaches and selecting the most suitable technology for each specific application.
Drawing on the academic papers presented, it is clear that it is essential to conduct a thorough comparison of \ac{NER} tools and select the optimal tool to meet specific requirements.
Apart from research studies that compare \ac{NER} tools, there is a noticeable lack of work that directly explores the perspectives of ML experts on the strategies and challenges involved in selecting and comparing \ac{NER} tools for specific use cases.
Amershi et al. \cite{amershi_software_2019} conducted a noteworthy survey at Microsoft, where they collected feedback from over 500 software engineers working on \ac{AI} and \ac{ML}. The survey's primary finding was that automation is crucial to facilitate efficient data aggregation, feature extraction, and label synthesis, thus accelerating the pace of experimentation.
Secondly, the survey found that \textit{``it is necessary to blend data management tools with their ML frameworks to avoid the fragmentation of data and model management activities''} \cite{amershi_software_2019}.
Finally, the survey \cite{amershi_software_2019} highlighted the importance of training and education for users with limited \ac{AI} experience, implying that a system that supports users in utilizing \ac{AI} could potentially reduce the need for such training.
The absence of research investigating the challenges faced by \ac{ML} experts in comparing \ac{NER} tools to select the appropriate solution for their project motivates the present study (\ac{RC}3). 

The authors identified three key \acp{RC} for conducting a survey among \ac{ML} experts regarding the challenges in comparing and selecting \ac{NER} tools. The first \ac{RC} addresses how \ac{ML} experts approach the decision-making process to compare \ac{NER} tools and choose the most appropriate ones. The second \ac{RC} focuses on the use of cloud resources for \ac{ML} model training for \ac{NER}. What are the key factors considered when deciding whether or which cloud-based resources to use? The third \ac{RC} highlights the lack of current research on the challenges that \ac{ML} experts face when comparing and selecting suitable \ac{NER} tools for their projects. After discussing the current state of the art in science and technology, identifying \acp{RC} in the areas of \ac{NER} and cloud, and reviewing related studies, the following chapter describes the modeling and design of the survey.

\section{Expert Survey Modeling}
\label{sec:theorybuilding}

\begin{figure}
 \centering
 \includegraphics[width=0.7\textwidth]{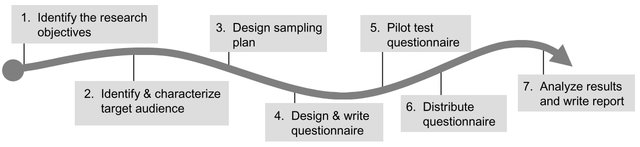}
 \caption[Survey Research Process by Kasunic]
 {Survey Research Process by Kasunic \cite{kasunic_designing_2005}}
 \label{fig:survey_research_process}
\end{figure}

This chapter focuses on the theory-building phase by addressing the \acp{RC} presented in Chapter \ref{sec:observation}, grounded in the current state of the art in science and technology. 
To address the \acp{RQ} defined in Chapter \ref{sec:introduction}, it is necessary to systematically collect and assess knowledge from experienced \ac{ML} experts in the field of \ac{NER}. Expert surveys are a widely used and effective method for obtaining opinions and insights from experts in a particular field \cite{auer_important_2021}. By gathering input from these experts, a deeper understanding of the topic can be gained and inform this research accordingly. This work aims to address the defined \acp{RQ} by surveying \ac{ML} experts in the field of \ac{NER} using the Kasunic model, an established framework for expert surveys \cite{kasunic_designing_2005}. The model outlines a comprehensive process consisting of seven stages to be followed when conducting a survey (see Figure \ref{fig:survey_research_process}). 
The stages outlined in the Kasunic model provide a structured approach to conducting expert surveys effectively, thereby ensuring that the resulting data is accurate and informative. This chapter will describe how to apply the guidelines and defined stages of the Kasunic model to survey \ac{ML} experts in the field of \ac{NER}. By following these guidelines and stages, meaningful experts' insights can be gathered and the \acp{RQ} can be addressed more effectively.

\textbf{\textit{Objectives.}} The first stage of the Kasunic model is to establish the \acp{RO} for the survey. As Chapter \ref{sec:introduction} has already defined the \acp{RQ} for this work, they can be used as a basis for establishing the \acp{RO}. The following objectives have been derived: \textit{(\ac{RO}1) Identify the criteria that \ac{ML} experts use to evaluate \ac{NER} tools and determine the importance of each criterion. (\ac{RO}2) Investigate the primary challenges that \ac{ML} experts encounter when selecting a suitable \ac{NER} tool.}

\textbf{\textit{Target Audience.}} Identifying and characterizing the target audience is the next stage in the survey process. The survey aims to gather insights from \ac{ML} experts with demonstrated experience in \ac{NER}. It is assumed that suitable experts should have more than three years of experience in \ac{ML}, \ac{NLP}, and \ac{NER}, and individuals who have written a Ph.D. thesis in any of these areas will also be considered. 

\textbf{\textit{Sampling Plan.}} In the third stage, the focus is on defining the sampling plan. This involves determining whether it is necessary to include a representative cross section of the target group in the survey and, if so, outlining how this can be accomplished.
The objective of this study is to gather information and challenges from experts using \ac{NER}, without the need to generalize the findings. Therefore, there is no requirement to cover a representative cross section of the target population or to develop a detailed sampling plan.
Participants were invited to participate on a voluntary basis in the survey. Initially, the survey included questions to collect information about the background and experiences of the participants. This approach will help ensure that the responses received come from individuals who meet the criteria to be considered expert in the field of \ac{NER}.

\textbf{\textit{Questionnaire.}} Following the design of the sampling plan, the next step was to create the questionnaire. It is crucial to carefully formulate the questionnaire items in a way that translates the \acp{RO}, as this facilitates the analysis and interpretation of the survey data. The questionnaire is structured into three sections: \textit{1. General Information}, \textit{2. Experience in selecting and using \ac{NER} Tools} and \textit{3. Final Questions}.
Section 1 is designed to gather essential demographic and background information about participants. It includes details such as their academic background, age, general experience with \ac{NER}, their specific roles in \ac{NER} projects, and the knowledge domains where \ac{NER} techniques have been applied.
At the end of the section, participants were asked about their experience with various \ac{NER} tools and frameworks.
The options included locally installable \ac{NER} tools and frameworks such as spaCy, Stanford Named Entity Recognizer, Hugging Face Transformers, Natural Language Toolkit, Flair, AllenNLP, OpenNLP, and GATE. Furthermore, frequently used cloud-based services such as IBM Watson Natural Language Understanding, Amazon Comprehend, Google Cloud Natural Language API, Microsoft Azure Cognitive Services, and OpenAI GPT-4 were considered to address the requirements of \ac{RC}2.
To compile the list of \ac{NER} tools and frameworks, a comprehensive review was carried out to identify commonly used \ac{NER} solutions.
In addition, participants could specify an additional tool in a free text field. Based on this input, further questions were posed in Section 2 regarding each selected tool or framework. These questions address the \acp{RC} 1-3 and \acp{RO} 1-2, including the comparison and evaluation of \ac{ML}-based \ac{NER} tools and challenges in tool selection. The questions for each framework in Section 2 are listed in Table \ref{table:tool_questionaire}.
Finally, Section 3 invites participants to rate their survey experience and includes an open-ended question, providing an opportunity for them to share additional thoughts or insights. The complete questionnaire is available online\footnote{\url{https://umfrage.fernuni-hagen.de/v3/671211?lang=en}}.

\begin{table}[!ht]
    \centering
    \caption{Questions per selected Tool}
    \label{table:tool_questionaire}
    \begin{tabularx}{\linewidth}{|>{\raggedright\arraybackslash}X|>{\raggedright\arraybackslash}X|>{\raggedright\arraybackslash}p{3.8cm}|}
        \hline
        \textbf{Question} & \textbf{Answer Options} & \textbf{Selection Type} \\
        \hline
    Please indicate your level of experience using $<selectedTool>$: & 1 (Very Poor), 2 (Poor), 3 (Average), 4 (Good), 5 (Excellent) & Single \\
        \hline
    How important were the following criteria in your evaluation of $<selectedTool>$ compared to other existing NER frameworks or tools, such $<examples>$? & Performance; Customization; Integration; Documentation and support; Licensing and cost; Accessibility; User interface and ease of use; Knowledge Domain Requirements; Privacy & Matrix, 5-point scale per option:\newline1 (Not Important),\newline2 (Slightly Important),\newline3 (Moderately Important),\newline4 (Important),\newline5 (Very Important) \\
        \hline
    In addition to the previously mentioned factors, were there any other criteria you considered important in your evaluation of $<selectedTool>$ compared to other existing NER frameworks or tools, such as $<examples>$? & $<openText>$ & Text \\
        \hline
    How important were the other criteria in your evaluation of $<selectedTool>$ compared to other existing NER frameworks or tools, such as $<examples>$? & 1 (Not Important), 2 (Slightly Important), 3 (Moderately Important), 4 (Important), 5 (Very Important) & Single \\
        \hline
    How hindering have the following challenges or limitations with $<selectedTool>$ been in the past? & Time and effort to learn the new framework; Lack of documentation; Challenges with integration into existing applications; Performance issues; Cost; Lack of support, such as documentation, community resources and paid support options & Matrix, 5-point scale per option\newline1 (Not Hindering),\newline2 (Slightly Hindering),\newline3 (Moderately Hindering),\newline4 (Hindering),\newline5 (Very Hindering) \\
        \hline
    Have you encountered any other challenges or limitations with $<selectedTool>$ in the past? If yes, please describe them briefly. & $<openText>$ & Text \\
        \hline
    How hindering have the other challenges or limitations with $<selectedTool>$ been in the past? & 1 (Not Hindering), 2 (Slightly Hindering), 3 (Moderately Hindering), 4 (Hindering), 5 (Very Hindering) & Single \\
        \hline
    \end{tabularx}
\end{table}

\textbf{\textit{Pilot Test Questionnaire.}} To eliminate errors and improve the questionnaire, it is important to test it with members of the target group. The test runs were conducted with a group of three \ac{ML} experts who have at least five years of experience in \ac{ML}, \ac{NLP}, and \ac{NER}. The survey was revised and improved based on the errors and problems identified during the testing process, which included the following: In an early draft of the questionnaire, respondents were asked which \ac{NER} tools they had experience with and which key factors were important to them when selecting \ac{NER} tools. During pretests, this approach was found to not allow for a specific identification of key factors that were important for the selection of a particular \ac{NER} tool. This is especially true when respondents had experience with multiple \ac{NER} tools. For example, performance might be a crucial factor when selecting a cloud-based tool, whereas privacy concerns might lead to the selection of locally installed tools.
As a result, the questionnaire was adjusted so that respondents could select from a list of \ac{NER} tools they had previously used. Then, specific questions were posed for each of the selected tools. However, further tests revealed that this approach increased the number of questions that needed to be answered in proportion to the number of \ac{NER} tools selected, making the questionnaire too lengthy overall. Consequently, the questions asked for each \ac{NER} tool were critically reviewed and reduced from 13 to 7.
Furthermore, it was found that the clarity of some questions could be improved by providing examples, which was subsequently implemented.

\textbf{\textit{Questionnaire Distribution.}} After completing quality assurance, the next step is to distribute the questionnaire to the appropriate target group. Chapter \ref{sec:implementation} provides a comprehensive description of the survey implementation.

\textbf{\textit{Analysis.}} Finally, the collected results were analyzed and presented using appropriate graphical representations to facilitate the understanding of the findings. The detailed report containing these diagrams is provided in the following chapter \ref{sec:evaluation}.

This chapter presented the modeling of the expert survey using a survey based on Kasunic's model \cite{kasunic_designing_2005}, encompassing all seven stages of the framework. The next chapter describes how the survey was conducted.

\section{Implementation}
\label{sec:implementation}
This chapter addresses the experimentation phase and provides a detailed description of how the survey was conducted, based on the modeling developed in the previous chapter.
The finalized questionnaire was distributed to selected members of the target group by email on August 19, 2024. Invitations were sent to a variety of individuals, including authors of relevant research papers, as well as employees of universities, research institutes, and industrial companies working in the field of \ac{ML} and \ac{NLP}. 
A total of 27 invitations were sent to individuals. In addition, an invitation was sent to an email distribution list of the Fraunhofer-Gesellschaft zur Förderung der angewandten Forschung e.V., which reaches approximately 60 researchers in the \ac{NLP} field. The survey was open for participation from August 18, 2024, to October 13, 2024, and received 23 responses, resulting in a response rate of 26\%. This response rate is better than that of other expert surveys such as \cite{auer_important_2021} (8\%), \cite{scheller_expert_2021} (11\%) or \cite{mendoza_relating_2019} (13\%).
After describing the conduct of the survey, the next chapter presents a summary and interpretation of the results of the expert survey.

\section{Evaluation}
\label{sec:evaluation}

In the previous chapter, the conduct of the expert survey was explained. This chapter focuses on the experimentation phase and presents a summary and interpretation of the results of the expert survey. It begins with a presentation of the background and demographics of the survey participants. The subsequent sections investigate the results of the expert survey based on the defined \acp{RO} 1-2. The sections cover a comparative analysis and evaluation of \ac{ML}-based \ac{NER} tools, and challenges encountered in the selection of tools. In addition, potential threats to the validity of the results are discussed.

\subsection{Demographics}

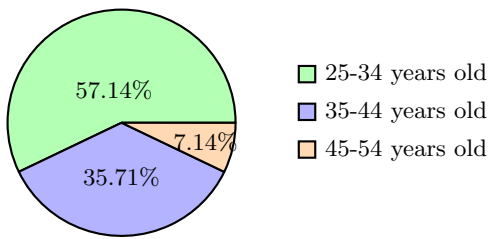
\begin{figure}[ht]
    \centering
    \begin{tikzpicture}[scale=0.75]
    \pie[text=legend, radius=2, color={green!30, blue!30, orange!30}]{
    57.14/25-34 years old,
    35.71/35-44 years old,
    7.14/45-54 years old
    }
    \end{tikzpicture}
    \caption{Age Distribution}
    \label{fig:age_distribution}
\end{figure}

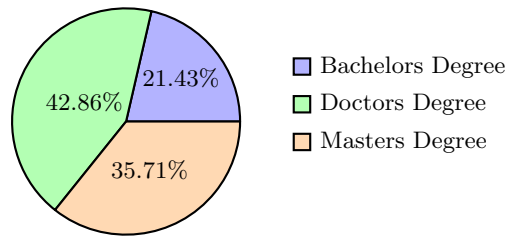
\begin{figure}[ht]
    \centering
    \begin{tikzpicture}[scale=0.75]
    \pie[text=legend, radius=2, color={blue!30, green!30, orange!30}]{
    21.43/Bachelors Degree,
    42.86/Doctors Degree,
    35.71/Masters Degree
    }
    \end{tikzpicture}
    \caption{Distribution of Academic Degrees}
    \label{fig:educational_attainment_distribution}
\end{figure}

\begin{figure}[ht]
    \centering
    \begin{tikzpicture}[scale=0.75]
    \pie[text=legend, radius=2, color={green!30, blue!30, orange!30}]{
    85.71/Computer Science,
    7.14/Economics,
    7.14/History
    }
    \end{tikzpicture}
    \caption{Field of Study Distribution}
    \label{fig:field_of_study_distribution}
\end{figure}
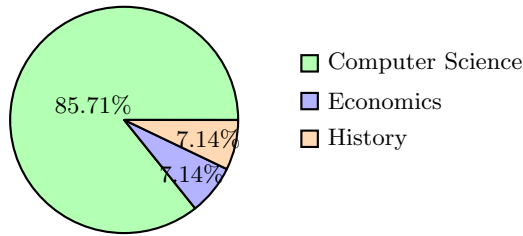

First, the demographic structure of the survey participants is analyzed. The participant group consisted predominantly of academics aged 25 to 34 years (57.14\%) and 35 to 44 years (35.71\%) (see Figure \ref{fig:age_distribution}). A smaller proportion of the participants was between 45 and 54 years old. As shown in Figure \ref{fig:educational_attainment_distribution}, most of the respondents had a doctorate (42. 86\%), while more than a third (35. 71\%) had a master's degree, and 21. 43\% had a bachelor's degree. This suggests that the survey was conducted with a highly qualified participant group. Most of the participants (85. 71\%) obtained their academic degrees in the field of Computer Science (Figure \ref{fig:field_of_study_distribution}). However, there were also participants from other disciplines, such as History and Economics. The aim of this work is to gain insight to support domain experts. Nevertheless, due to the predominant Computer Science backgrounds of the participants, caution is warranted. The results of this survey cannot be directly generalized to domain experts and must be interpreted and evaluated accordingly.

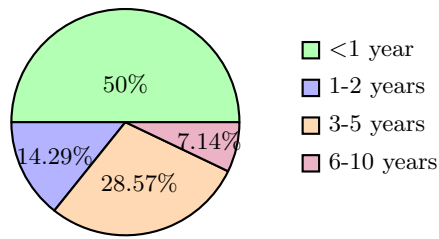
\begin{figure}[ht]
    \centering
    \begin{tikzpicture}[scale=0.75]
    \pie[text=legend, radius=2, color={green!30, blue!30, orange!30, purple!30, yellow!30}]{
    50/\textless 1 year,
    14.29/1-2 years,
    28.57/3-5 years,
    7.14/6-10 years
    }
    \end{tikzpicture}
    \caption{Distribution of NER Experience by Years}
    \label{fig:exeperience_distribution}
\end{figure}

\begin{figure}[ht]
    \centering
    \begin{tikzpicture}[scale=0.75]
    \pie[text=legend, radius=2, color={blue!30, green!30, orange!30, purple!30, yellow!30}]{
    35.71/2 (Poor),
    42.86/3 (Average),
    21.43/4 (Good)
    }
    \end{tikzpicture}
    \caption{Distribution of NER Experience Level}
    \label{fig:exeperience_level_distribution}
\end{figure}
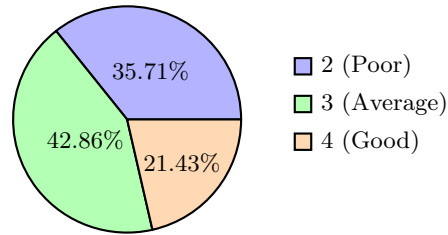

Now, the experiences of the participants in the field of \ac{NER} was examined. Figure \ref{fig:exeperience_distribution} illustrates that 50\% of the participants have been involved with \ac{NER} for less than a year, while 14.20\% have been involved for up to two years. In contrast, 28.57\% of the participants have accumulated over three years of experience in this area. Only 35.71\% of the respondents reported having minimal experience with \ac{NER} (Figure \ref{fig:exeperience_level_distribution}). The majority rate their \ac{NER} experience as average (42.86\%) or good (21.43\%). In summary, given the high level of education of the participants, this group can be considered experienced in \ac{NER}, even though many have only recently entered this field.

\begin{figure}[ht]
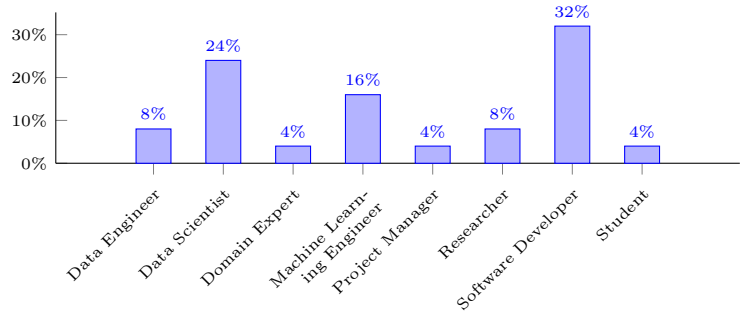

	\centering
    \newybarchartpct{roles.csv}{}
	\caption{Distribution of Roles in NER Projects}
	\label{fig:role_distribution}
\end{figure}

\begin{figure}[ht!]
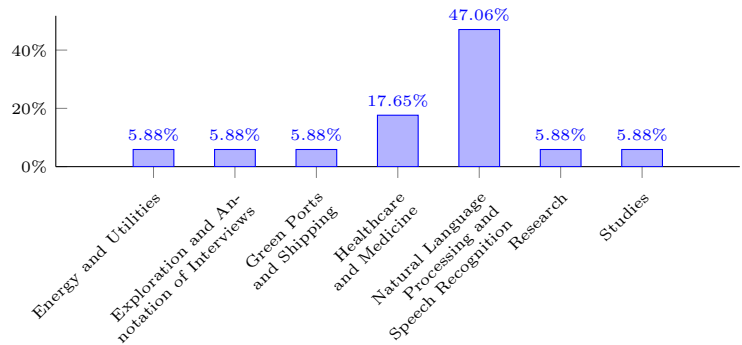

	\centering
    \newybarchartpct{domains.csv}{}
	\caption{Distribution of NER Project Domains}
	\label{fig:domain_distribution}
\end{figure}

Although most of the participants come from the field of Computer Science, they apply \ac{NER} in their projects across a variety of roles and domains. The question \textit{``In your current or most recent Named Entity Recognition project, what was your primary responsibility or role?''} allowed multiple selections. This led to a diverse representation of roles in which participants were involved in their \ac{NER} projects, such as Software Developer (32\%), Data Scientist (24\%) or Machine Learning Engineer (16\%), as well as Domain Expert or Project Manager (each 4\%) (Figure \ref{fig:role_distribution}).
As indicated in Figure \ref{fig:domain_distribution} for the question \textit{``In which domains have you previously applied Named Entity Recognition techniques?''}, \ac{NER} was primarily utilized in the field of \ac{NLP} and speech recognition (47.06\%). In addition, a manifold picture emerges: \ac{NER} was also applied in domains such as Healthcare and Medicine (17.65\%), as well as Green Ports and Shipping and E-Commerce and Retail (each 5.88\%). Again, multiple selections were allowed.

\subsection{Experience with Selecting and using NER Tools}

\begin{figure}[ht]
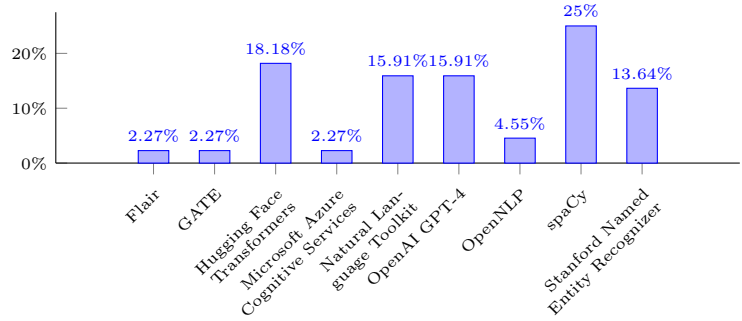

	\centering
    \newybarchartpct{frameworks.csv}{}
	\caption{Distribution of Experience with NER Framework}
	\label{fig:framework_distribution}
\end{figure}

After initially analyzing the demographic structure of the participants, the second step provides a detailed examination of the \ac{NER} frameworks employed. For this purpose, participants were asked which \ac{NER} tools and frameworks they have experience with, allowing multiple selections. The results are presented in Figure 9. The most frequently mentioned \ac{NER} tools and frameworks were spaCy (25\%), Hugging Face Transformers (18.18\%), OpenAI GPT-4 (15.91\%), Natural Language Toolkit (15.91\%), and Stanford Named Entity Recognizer (13.64\%). All other systems remained below 5\% or were not mentioned at all.


\begin{figure}[ht]
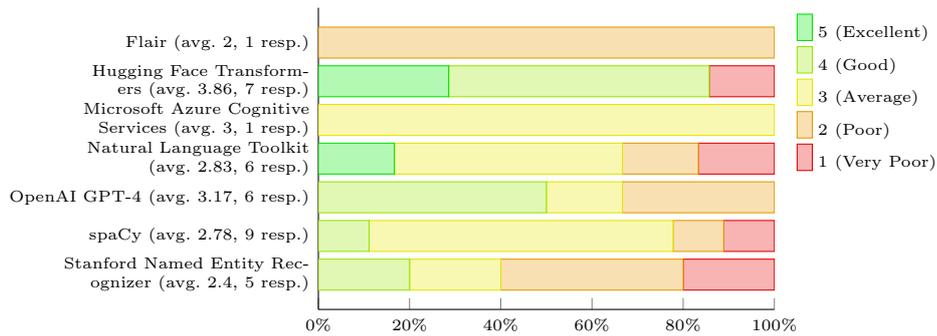

	\centering
    \newxbarchartstackedpct{frameworks_experience_level.csv}{}
	\caption{Distribution of Experience Levels per NER Framework}
	\label{fig:frameworks_experience_level}
\end{figure}

The levels of experience of participants with various \ac{NER} frameworks (Figure \ref{fig:frameworks_experience_level}) reveal that Hugging Face Transformers (average 3.86 from 7 responses) and OpenAI GPT-4 (average 3.17 from 6 responses) achieved the highest ratings. These tools were rated favorably for ease of use and effectiveness. The Hugging Face Transformers library, widely regarded for its intuitive application of \acp{LLM} \cite{jain_introduction_2022}, appears to facilitate rapid learning and adoption, making it a preferred choice for both novices and experienced developers alike. OpenAI GPT-4, while designed primarily for text generation, is highly adaptable for \ac{NER} tasks due to its user-friendly natural language prompting mechanism \cite{openaiGPT4TechnicalReport2024}.
Other frameworks, such as Microsoft Azure Cognitive Services (average 3.1, 1 response) and Natural Language Toolkit (average 2.83, 6 responses), showed more modest ratings. Flair, spaCy, and Stanford Named Entity Recognizer scored lower in terms of perceived usability (averages ranging between 2.4 and 2.78), likely reflecting their steeper learning curves or limitations in broader applicability.


\begin{figure}[ht]
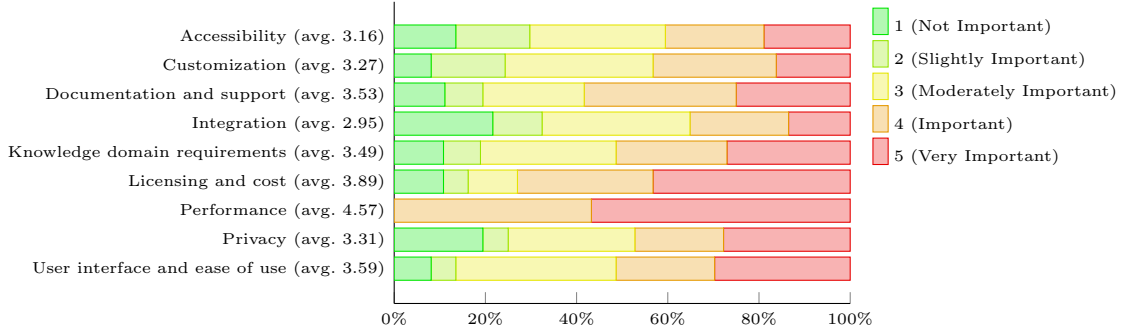

	\centering
    \newxbarchartstackedpct{average_priority.csv}{}
	\caption{Priority Distribution per Selection Criteria}
	\label{fig:average_priority}
\end{figure}

Figure \ref{fig:average_priority} details the evaluations of participants of various selection criteria across all \ac{NER} tools and frameworks, revealing performance (average 4.57) as the most critical factor. The consistently high prioritization of performance underscores the need for \ac{NER} tools to deliver accurate and reliable results in different operational contexts.
As described in Chapter \ref{sec:theorybuilding}, questions were also posed regarding cloud-based tools, such as Microsoft Azure Cognitive Services. Especially, for these cloud-based tools, licensing and cost (average 3.89), as well as user interface and ease of use (average 3.59), also rank highly.
This reflects the growing importance of affordability and accessibility in encouraging adoption among diverse user groups, such as newbies and \ac{ML} experts.
In contrast, factors such as integration (average 2.95) and customization (average 3.27) were moderately important, suggesting that participants found these areas less critical when evaluating \ac{NER} tools. Privacy (average 3.31) received mixed scores, signaling varying levels of concern depending on the application and deployment model.
Participants also had the opportunity to specify and evaluate additional criteria in a free text field. For spaCy ``GPU compatibility for Transformers'' was mentioned and rated as ``3 (Moderately Important)''. For OpenAI GPT-4, ``Response Time'' was noted and rated as ``5 (Very Important)'', which can be classified as performance.
In general, it can be concluded that all the criteria were considered relevant. This suggests that the importance of criteria is highly dependent on the specific project and that there are no criteria that can be universally deemed unimportant. 
This is also illustrated by the fact that there are different specific challenges for \ac{NER}, depending on which knowledge domain \ac{NER} is to be applied \cite{pakhale_comprehensive_2023}.
However, the performance criterion remains significant in all cases.
The significance of performance in \ac{NER} tools and frameworks is underscored by numerous studies that compare their effectiveness \cite{schmitt_replicable_2019, casey_systematic_2021, weiying_benchmarking_2019}. When analyzing the results of cloud-based and locally installable \ac{NER} tools and frameworks, as outlined in Chapter \ref{sec:theorybuilding}, and calculating the average performance values, notable insights emerge. Table \ref{table:criteria_cloud_vs_local} illustrates the average results for both types of tools, highlighting the differences (Delta) between them.
For locally installable systems, documentation and support play a critical role, as evidenced by a Delta of -1.14. This observation aligns with findings from related studies, where documentation is frequently emphasized as a key factor in evaluating \ac{NER} tools. For instance, Schmitt et al. argue that criteria such as documentation should be carefully assessed before selecting and deploying an \ac{NER} solution \cite{schmitt_replicable_2019}.
However, for cloud-based systems, the user interface and ease of use are particularly relevant (Delta 1.09). Tamla et al. have already pointed out that managing cloud-based resources is a challenge for newbies and \ac{ML} experts \cite{tamla_cie_2023}, thus increasing the relevance of the user interface and usability for the use of cloud-based resources. Kurdi et al. also recognized that a good user interface is important for the usability of cloud-based services \cite{hutchison_towards_2014}. Thus, it is clear that users of cloud-based \ac{NER} services could benefit from systems that simplify their use.

\begin{table}[ht]
\centering
    \caption{Comparison of Average Priority per Selection Criteria}
    \label{table:criteria_cloud_vs_local}
    \begin{tabular}{| l  |l  |l |l|}
        \hline
        \textbf{Criteria} & \textbf{Cloud} & \textbf{Local}  &\textbf{Delta}\\
        \hline
        Accessibility &  3.57 &  3.0  &0.57
\\
        Customization &  3.43 &  3.31  &0.12
\\
        Documentation and support &  2.57 &  3.71  &\cellcolor{yellow}-1.14
\\
        Integration &  2.57 &  3.1  &-0.53
\\
        Knowledge Domain Requirements &  3.43 &  3.45  &-0.02
\\
        Licensing and cost &  3.71 &  3.9  &-0.19
\\
        Performance &  4.57 &  4.55  &0.02
\\
        Privacy &  3.14 &  3.29  &-0.15
\\
        User interface and ease of use &  4.43 &  3.34  &\cellcolor{yellow}1.09\\
        \hline
    \end{tabular}
\end{table}


\begin{figure}[ht]
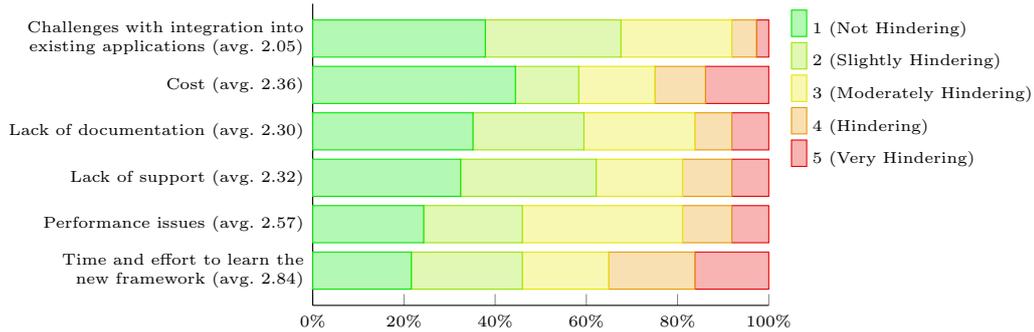

	\centering
    \newxbarchartstackedpct{average_hindering.csv}{}
	\caption{Priority Distribution per Hindrance Criteria}
	\label{fig:average_hindering}
\end{figure}

\begin{table}[ht]
\centering
    \caption{Comparison of Average Priority per Hindrance Criteria}
    \label{table:hindrance_cloud_vs_local}
    \begin{tabular}{| l  |l  |l |l|}
        \hline
        \textbf{Hindrance} & \textbf{Cloud} & \textbf{Local}  & \textbf{Delta}
\\
        \hline
        Challenges with integration into existing applications & 2.29 & 1.90  &0.39
\\
        Cost & 3.71 & 1.93  &\cellcolor{yellow}1.79
\\
        Lack of documentation & 2.43 & 2.31  &0.12
\\
        Lack of support & 3.00 & 2.21  &0.79
\\
        Performance issues & 2.29 & 2.62  &-0.33
\\
        Time and effort to learn the new framework & 2.43 & 3.00  &\cellcolor{yellow}-0.57\\
        \hline
    \end{tabular}
\end{table}

The responses to the question \textit{ ``How hindering have the following challenges or limitations with $<selectedTool>$ been in the past?''} for all \ac{NER} tools and frameworks are presented in Figure \ref{fig:average_hindering}. The responses are surprising in that very few challenges were classified by participants as very hindering. The most frequently cited issue was ``Time and effort to learn the new framework'' (average 2.84). This was followed by ``Performance Issues''  (average 2.57), which is consistent with the findings in Figure \ref{fig:average_priority}. ``Challenges with integration into existing applications'' were mentioned the least as a hindrance (average 2.05). Other challenges, such as ``Cost'' (average 2.36), ``Lack of support'' (average 2.32), and ``Lack of documentation'' (average 2.30), were ranked mid-range but low. Each challenge was mentioned at least once as hindering or very hindering, supporting the assertion that the requirements for \ac{NER} tools and frameworks are project specific. 
For spaCy, it was also noted that there are difficulties with supported file formats: \textit{``limited compatibility with common annotation formats; must be converted into spaCy’s own format''}. This was rated as ``3 (Moderately Hindering''. For OpenAI GPT-4, the following challenges were mentioned in the free text field and rated as ``4 (Hindering)'': \textit{``licensing, privacy considerations, and closed source''}.
In general, it can be concluded that reducing the time and effort required to learn new frameworks is essential. Here, also, the results of this question were grouped with respect to cloud-based and locally installable \ac{NER} tools and frameworks and the average value was calculated. As shown in Table \ref{table:hindrance_cloud_vs_local}, the time and effort required to learn the new framework are particularly restrictive for locally installable systems (Delta -0.57). 
Therefore, a system that supports the comparison and selection of \ac{NER} tools and frameworks must also assist users to work quickly and easily with the various systems. For cloud-based services, costs are, as expected, a particularly significant barrier to their adoption (Delta 1.79). Therefore, when using cloud-based resources, it is important to control costs and pay attention to cost efficiency \cite{chard_cost-aware_2015}.

In addition to the specified \ac{NER} tools and frameworks, the participants were able to indicate experience with other \ac{NER} tools or frameworks. One participant noted the use of \textit{``Self-hosted open source LLM (many)''}, suggesting that he downloaded freely available \acp{LLM} from the Internet, such as those from Hugging Face\footnote{\url{https://huggingface.co/models}}, for local \ac{NER} applications. This participant rated their expertise with this technology as ``5 (Very High)''.
When asked, \textit{``How important were the following criteria in your evaluation of your selected NER tool or framework compared to other existing NER frameworks or tools, such as spaCy or Amazon Comprehend?''} the participant rated the criteria: Documentation and support, licensing and cost, accessibility, user interface and ease of use, knowledge domain requirements, and privacy as ``5 (Very Important)''. Customization and integration were rated only ``1 (Not Important)''.
Regarding the challenges or limitations of this technology, the challenges with integration into existing applications and cost were rated ``5 (Very Hindering)'', while performance issues received a ``3 (Moderately Hindering)''. Other factors, such as time and effort to learn the new framework, lack of documentation, and lack of support, such as documentation, community resources, and paid support options, were rated ``1 (Not Hindering)''.
This suggests that the respondent has high expectations for an \ac{NER} tool or framework, leading to the use of local open-source \acp{LLM} for \ac{NER}, while considering him an expert in this area. However, it remains unclear which tools and frameworks were used alongside the local \acp{LLM}. It is likely that a system from the specified selection was employed, such as Hugging Face Transformers.
The rationale behind the high ratings for integration challenges and costs remains ambiguous. 
It is possible that the question was misinterpreted and answered in the context of other solutions, given that open-source \acp{LLM} typically offer low costs and high flexibility \cite{yang_fingpt_2023}. Should the responses have indeed pertained to open-source \acp{LLM}, the effort involved in utilizing local open-source \acp{LLM} may have resulted in increased costs and challenges related to integration.
Ultimately, a significant finding of this study is that locally operated open-source \acp{LLM} represent a relevant technology for \ac{NER} and are already being used in other projects \cite{yang_fingpt_2023}.
Open-source \acp{LLM} must be considered in the comparison and selection of suitable \ac{NER} tools and frameworks.

\subsection{Final Questions}

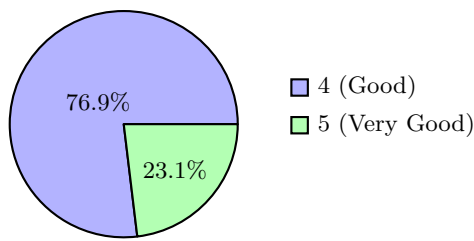
\begin{figure}[ht]
    \centering
    \begin{tikzpicture}[scale=0.75]
    \pie[text=legend, radius=2, color={blue!30, green!30, orange!30, purple!30, yellow!30}]{
    76.9/4 (Good),
    23.1/5 (Very Good)
    }
    \end{tikzpicture}
    \caption{Distribution of Survey Experience}
    \label{fig:exeperience_survey}
\end{figure}

Finally, participants were asked to assess their experience with the survey and to provide any additional comments. The survey received overwhelmingly positive feedback, as illustrated in Figure \ref{fig:exeperience_survey}.
In the final remarks, it was noted that some questions were occasionally too detailed. This feedback can be incorporated into future surveys, although careful consideration must be given to whether such detailed questions are necessary for achieving the survey's objectives.

\subsection{Threats to Validity}

When conducting this survey, several potential threats to the validity of the results should be considered.
First, the 23-part participant count can be deemed too low, which could limit the generalizability of the findings. However, the survey primarily aims to provide information on the relevant criteria to compare and select \ac{NER} tools and frameworks. This potentially limited generalizability thus poses a minor risk.
Second, most of the participants come from the field of computer science. This one-sided composition may lead to biases, as the perspectives and experiences of this group do not necessarily reflect the views of other relevant disciplines. While the responses from computer scientists are significant for domain experts, they should be carefully translated to meet the specific needs of those experts. Future work may, therefore, consider surveying additional domain experts.
Third, a participant noted that some questions were too detailed. This may have resulted in some participants not completing the survey in full or having difficulty answering the questions appropriately. However, it was important to ask specific questions to obtain precise answers. In future surveys, the level of detail in the questions should be critically reassessed.
The mentioned threats to validity have been adequately considered in interpreting the results. Consequently, the remaining risk to the findings of this survey is assessed as low.

\section{Conclusion}
\label{sec:conclusion}
In this article, an expert-based survey was presented to gain insights into how \ac{NER} experts compare and select \ac{NER} tools and frameworks, as well as the challenges they face. The structured research methodology proposed by Nunamaker was applied. Chapter \ref{sec:introduction} introduces the research area and motivates the work through related research projects. Chapter \ref{sec:observation} analyzes the current state of science and technology, reviews and compares similar works, and presents the \acp{RC}. In Chapter \ref{sec:theorybuilding}, the \acp{RO} were defined on the basis of Kasunic's Survey Research Process, followed by the design and quality assurance of the survey. The survey publication is described in Chapter \ref{sec:implementation}. Finally, Chapter \ref{sec:evaluation} provides a detailed analysis and interpretation of the results.
Various results were obtained related to \ac{RO}1, which focuses on identifying relevant selection criteria for \ac{NER} tools and frameworks. The survey highlighted that performance is a particularly important criterion. Furthermore, expert opinions varied significantly. All specified selection criteria were regarded important or very important at least once. This indicates that the relevance of the criteria may vary depending on the project. Therefore, a supportive system should be flexible enough to accommodate various criteria along with performance.
In the context of \ac{RO}2, a distinction is made between cloud-based services and locally installed tools and frameworks. For cloud-based services, cost and user-friendliness are particularly significant. Both aspects represent important requirements for a system that uses cloud-based services for \ac{NER}. In the case of locally installed systems, the effort required for users to adopt a new system should be minimized, which can be facilitated by an appropriate software solution. Additionally, there were indications of the relevance of locally installed open-source large language models, which should also be integrated into future software systems.

In summary, the defined research objectives have been successfully achieved, and the \acp{RC} identified in Chapter \ref{sec:observation} have been addressed. The results provide valuable information for future systems designed to support the selection and comparison of \ac{NER} tools and frameworks. Future work should focus on adapting these results to the needs of domain experts, enabling them to utilize \ac{NER} in the development of \acp{CPG}.

\bibliography{paper}





\end{document}

%% file: acronyms.tex
\begin{acronym}

\acro{AC}{Affective Computing}
\acro{AI}{Artificial Intelligence}
\acro{AI4H3}{Artificial Intelligence for Hospitals, Healthcare \& Humanity}
\acro{API}{Application Programming Interface}
\acro{AWS}{Amazon Web Service}
\acro{BERT}{Bidirectional Encoder Representations from Transformers}
\acro{CDS}{Clinical Decision Support}
\acro{CDSS}{Clinical Decision Support System}
\acro{CIE}{Cloud-based Information Extraction}
\acro{CKMS}{Content and Knowledge Management Subsystem}
\acro{CPG}{Clinical Practice Guideline}
\acro{depparse}{Dependency Parsing}
\acro{DSS}{Decision Support System}
\acro{EHR}{Electronic Health Record}
\acro{ELMo}{Embeddings from Language Models}
\acro{eNE}{emerging Named Entity}
\acro{EU}{European Union}
\acro{F1}{F-Score}
\acro{FIT4NER}{Framework-Independent Toolkit for Named Entity Recognition}
\acro{FTK}{FTK e.V. Research Institute for Telecommunications and Cooperation}
\acro{GPT}{Generative Pre-trained Transformer}
\acro{GUI}{Graphical User Interface}
\acro{H2020}{Horizon 2020}
\acro{HPC}{High-Performance Computing}
\acro{IaaS}{Infrastructure as a Service}
\acro{IE}{Information Extraction}
\acro{IO}{Information Overload}
\acro{IOT}{Internet of Things}
\acro{IR}{Information Retrieval}
\acro{IS}{Information System}
\acro{K8S}{Kubernetes}
\acro{KlinSH-EP}{KlinGard Smart Hospital EcoSystem Portal}
\acro{KM-EP}{Content and Knowledge Management Ecosystem Portal}
\acro{KMS}{Knowledge Management System}
\acro{LLM}{Large Language Model}
\acro{MES}{Medical Expert System}
\acro{ML}{Machine Learning}
\acro{MVC}{Model-View-Controller}
\acro{NE}{Named Entity}
\acro{NER}{Named Entity Recognition}
\acro{NLP}{Natural Language Processing}
\acro{P}{Precision}
\acro{PaaS}{Platform as a Service}
\acro{PS}{Problem Statement}
\acro{R}{Recall}
\acro{RAGE}{Realising an Applied Gaming Ecosystem}
\acro{RBES}{Rule-based Expert System}
\acro{RC}{Remaining Challenge}
\acro{RecomRatio}{Recommendation Rationalisation}
\acro{RO}{Research Objective}
\acro{RQ}{Research Question}
\acro{SaaS}{Software as a Service}
\acro{SG}{Serious Game}
\acro{Smart CDSS}{Smart Clinical Decision Support System}
\acro{SNERC}{Stanford Named Entity Recognition and Classification}
\acro{SO}{Survey Objective}
\acro{UC}{Use Case}
\acro{UCSD}{User Centered System Design}
\acro{WWW}{World Wide Web}
\acroplural{AWS}[AWS]{Amazon Web Services}
\acroplural{eNE}[eNEs]{emerging Named Entities}
\acroplural{NE}[NEs]{Named Entities}
\end{acronym}